\documentclass{ws-procs975x65}

\def\beq{\begin{equation}}
\def\eeq{\end{equation}}
\def\bea{\begin{eqnarray}}
\def\eea{\end{eqnarray}}
\def\bq{\begin{quote}}

\def\frac#1#2{{\textstyle{{#1}\over {#2}}}}

\def\lsim{\mathrel{\rlap{\lower4pt\hbox{\hskip1pt$\sim$}}
    \raise1pt\hbox{$<$}}}
\def\gsim{\mathrel{\rlap{\lower4pt\hbox{\hskip1pt$\sim$}}
    \raise1pt\hbox{$>$}}}
\def\sqr#1#2{{\vcenter{\vbox{\hrule height.#2pt
         \hbox{\vrule width.#2pt height#1pt \kern#1pt
         \vrule width.#2pt}
         \hrule height.#2pt}}}}

\def\AJ{{\it Ap. J.} }
\def\AJL{{\it Ap. J. Lett.} }

\def\ASAS{{\it Astron. and Astrophys.} }

\def\JP{{\it J. Phys.} }

\def\NP{{\it Nucl. Phys.} }
\def\PL{{\it Phys. Lett.} }
\def\PR{{\it Phys. Rev.} }
\def\PRL{{\it Phys. Rev. Lett.} }

\def\gappeq{\mathrel{\rlap {\raise.5ex\hbox{$>$}}
{\lower.5ex\hbox{$\sim$}}}}
 
\def\lappeq{\mathrel{\rlap{\raise.5ex\hbox{$<$}}
{\lower.5ex\hbox{$\sim$}}}}

\begin{document}

\title{Translational Invariance and Noncommutative Field Theories}

\author{Orfeu Bertolami}

\address{Instituto Superior T\'ecnico, Departamento de F\'\i sica, 
Av. Rovisco Pais 1, 1049-001 Lisboa, Portugal\\E-mail: orfeu@cosmos.ist.utl.pt}


\maketitle

\abstracts{
Implications of noncommutative field theories with commutator of the coordinates of
the form $\left[x^{\mu },x^{\nu }\right]=i\, \Lambda _{\quad \omega }^{\mu \nu }x^{\omega }$
with nilpotent structure constants are investigated. It is shown that a free quantum field
theory is not affected by noncommutativity, but that invariance under translations is broken and 
hence the energy-momentum conservation is not respected. The new energy-momentum law
is expressed by a Poincar\'e-invariant equation and the resulting
kinematics is developed and applied to the astrophysical
puzzle related with the observed violation of the GZK cutoff.}

\vspace{0.15cm}

\noindent
\centerline{{\bf Dedicated to the memory of Lu\'\i s Guisado}}

\vspace{0.15cm} 

\section{Introduction}

Achieving a consistent theory of quantum gravity is one of the major goals of XXI century physics. String/M-theory 
is the best candidate so far for this synthesis:

\vspace{0.15cm}

\noindent
1. It is a finite quantum theory for gravity and contains, thanks to the 
mechanism of cancellation of gauge and gravitational anomalies, grand unified theories 
in a constrained way.

\vspace{0.15cm}

\noindent
2. It is naturally supersymmetric, and hence leads to models that are free from the hierarchy problem. 

\vspace{0.15cm}

\noindent
3. It has black hole solutions, and therefore allows addressing key issues associated with those singular 
objects, such as the physics underlying  
their entropy and their presumed non-unitary evolution.

\vspace{0.15cm}

\noindent
4. It is a natural framework for many important ideas and techniques in the 
field theory such as Supergravity, Kaluza-Klein mechanism, 
conformal field theory, noncommutative geometry, braneworld scenarios, etc. 

\vspace{0.15cm}

Despite these appealing features, string theory has so far neither provided a decisive insight toward a solution of 
the cosmological constant problem\cite{Witten} nor has unambiguously suggested a clear cut set of phenomenological signatures. 
This last task would be simpler if 
among the predictions of string theory were the breaking of fundamental symmetries such as Lorentz 
invariance, CPT symmetry and the Equivalence Principle in phenomenologically testable ranges. 
It is remarkable that the spontaneous breaking of Lorentz and CPT 
symmetries can occur in string field theory\cite{Kostelecky1,Kostelecky2}, but phenomenological 
implications are shown the be fairly subtle, probably suppressed by powers of the Planck mass. 
The same can be stated about the Equivalence Principle\cite{Damour}. 

Given these difficulties, it is particularly relevant to broaden the scope of the search for 
experimental evidences of quantum gravity phenomena. Even though it is not beyond dispute, 
it has been argued that evidence for the breaking of Lorentz symmetry may have already been encountered 
in cosmic ray physics and high-energy astrophysics:

\vskip 0.15cm

\noindent
1. In the observation of ultra-high energy cosmic rays\cite{Hayashida,Bird,Lawrence,Efimov}  
with energies
beyond the Greisen-Zatsepin-Kuzmin (GZK) 
cutoff, $E_{GZK} \simeq 4 \times 10^{19}~eV$ \cite{Greisen}, even though, the compatibility of these observations 
with the preliminary measurements of the energy spectrum by the HiRes 
Collaboration is still under debate\cite{Abbasi}. These events, are a challenge to present knowledge   
to accelerate cosmic particles and may require the violation of Lorentz invariance as an explanation 
whether their sources are shown to lie beyond $50 - 100~Mpc$ \cite{Stecker}. This arises as breaking of Lorentz symmetry  
suppresses resonant scattering reactions of the primaries with photons of the Cosmic Microwave 
Background (CMB)\cite{Sato1,Coleman,Mestres,Bertolami1}. We mention that no conclusive correlation with astrophysical sources 
has ever been found\cite{Bertolami2}.

\vskip 0.15cm

\noindent
2. In the observation of gamma radiation from distant sources such as Markarian 421 and Markarian 501 blazars 
with energies above $20~TeV$ \cite{Krennrich}. These observations 
suggest a violation of Lorentz symmetry as otherwise 
there should exist a strong attenuation of fluxes beyond $100~Mpc$ 
of $\gamma$-rays with energies higher 
than $10~TeV$ by the diffuse extragalactic background of infrared photons due to pair creation\cite{Amelino1}.

\vskip 0.15cm

\noindent
3. In the evolution of air showers produced by ultra high-energy hadronic particles which  
suggests that pions are more stable than predicted\cite{Antonov}. 

\vskip 0.2cm

As already mentioned, Lorentz invariance can be spontaneously broken due to
non-trivial solutions in string field theory\cite{Kostelecky1}, but may also arise 
in loop quantum gravity\cite{Gambini,Alfaro}, 
and in quantum gravity inspired spacetime foam scenarios\cite{Aloisio}. 
A violation of the Lorentz symmetry may also lead to
the breaking of CPT symmetry\cite{Kostelecky2}.
There exists a workable extension of the Standard Model inspired in string field theory 
that incorporates violations 
of Lorentz and CPT symmetries\cite{Colladay}. In the context of this extension 
several questions can be addressed, such as 
the violation the of GZK cutoff\cite{Bertolami1}, the generation of the  
baryon asymmetry of the Universe\cite{Bertolami4} and of primordial magnetic fields\cite{Bertolami5}.
The breaking of Lorentz symmetry also arises
in models where the electromagnetic coupling evolves\cite{Lehnert1,Bertolami6} 
and may also be related to the vacuum energy density\cite{Bertolami7}. Several other consequences to 
particle physics have already been worked out\cite{Kostelecky3}. 

Another important class of theories where Lorentz invariance is not respect are
noncommutative field theories\cite{Carroll}. Noncommutative theories 
have been extensively studied given that they naturally emerge in string theory\cite{Seiberg_Witten},
and also due to their interesting properties and implications for
field theory\cite{Doug_Nekra}. In what concerns the coupling with gravity in noncommutative theories, 
a noncommutative scalar field theory has already been examined\cite{Bertolami8}.

In what follows we shall consider classical and quantum field theory features of models where the noncommutativity
of the coordinates has the following form

\begin{equation}
\left[x^{\mu },x^{\nu }\right]=i\, \Lambda _{\quad \omega }^{\mu \nu }x^{\omega }~,
\label{intro}
\end{equation}
together with the condition of nil-potency specified below.
This leads to a violation of the symmetry under translations and,
consequently, requires a reformulation of the energy-momentum conservation\cite{Bertolami9}. 
This work has been developed 
in collaboration with Lu\'\i s Guisado, who tragically died in a car crash 
on June 28th, 2003. Lu\'\i s was a bright 23 years old graduate student and was regarded as one of 
the great hopes 
of portuguese theoretical physics. I would like to dedicate this contribution to the memory of his kind 
and friendly person.

\section{Mathematical Considerations}

%\subsection{Noncommutative Algebra}

A noncommutative and associative product can be defined through the Lie-algebra
commutator Eq. (\ref{intro}), where $\Lambda ^{\mu \nu \omega }$ is 
a real tensor with units of $\textrm{mass}^{-1}$ and $\Lambda ^{\mu \nu \omega }=-\Lambda ^{\nu \mu \omega }$. 
Associativity implies in the Jacobi identity

\begin{equation}
\Lambda _{\quad \omega }^{\mu \nu }\Lambda _{\quad \beta }^{\omega \alpha }+
\Lambda _{\quad \omega }^{\nu \alpha }\Lambda _{\quad \beta }^{\omega \mu }+
\Lambda _{\quad \omega }^{\alpha \mu }\Lambda _{\quad \beta }^{\omega \nu }=0\: .
\label{Jacobi}
\end{equation}

A noncommutative Fourier mode can be defined by

\begin{equation}
e_{*}^{ik\cdot x}=\sum _{n=0}^{\infty }{i^{n} \over n!}\overbrace{\left(k\cdot x\right)*...*
\left(k\cdot x\right)}^{n\textrm{ factors}}=\sum _{n=0}^{\infty }{i^{n} \over n!}\left(k\cdot x\right)_{*}^{n}~~,
\label{Fourier}
\end{equation}
so that the functional space spanned by these Fourier modes,
with elements of the form

\begin{equation}
f\left(x\right)=\int {d^{n}k \over \left(2\pi \right)^{n}}\tilde{f}\left(k\right)e_{*}^{ik\cdot x}
\label{NC_Fourier}
\end{equation}
reduces, in the commutative limit, to the usual Hilbert space. 

The product of two generic functions is given by

\begin{equation}
f*g=\int {d^{n}k \over \left(2\pi \right)^{n}} 
{d^{n}q \over \left(2\pi \right)^{n}}\tilde{f}\left(k\right)\tilde{g}\left(q\right)e_{*}^{ik\cdot x}*e_{*}^{iq\cdot x},
\end{equation}
where the functions are expressed in terms of their noncommutative
Fourier expansion. This product is completely determined if the product
of two Fourier modes $e_{*}^{ik\cdot x}*e_{*}^{iq\cdot x}$ can be
evaluated. This is achieved through the Baker-Hausdorff
formula 

\begin{equation}
e_{*}^{ik\cdot x}*e_{*}^{iq\cdot x}=\exp _{*}\left\{ i\left(k+q\right)\cdot x+{1 \over 2}\left[ik\cdot x,iq\cdot x\right]+...\right\} ,
\end{equation}
where the dots stand for higher order commutators. Since the commutators
obey

\begin{equation}
\left[x^{\mu _{1}},\left[x^{\mu _{2}},...,\left[x^{\mu _{n}},x^{\nu }\right]\right]...\right]\propto i^{n}x^{\omega }\; ,
\end{equation}
the product of two Fourier modes is a Fourier mode

\begin{equation}
e_{*}^{ik\cdot x}*e_{*}^{iq\cdot x}=e_{*}^{i\left[k+q+V\left(k,q\right)\right]\cdot x}
\end{equation}
with $V$ being determined by the Baker-Hausdorff expansion:

\begin{equation}
V_{\omega }\left(k,q\right)=k_{\mu }q_{\nu }\Lambda _{\quad \lambda }^{\mu \nu }\left[-{1 \over 2}\delta _{\omega }^{\lambda }
+{k_{\alpha }-q_{\alpha } \over 12}\Lambda _{\quad \omega }^{\alpha \lambda }\right]+{O\left(\Lambda ^{3}\right)}\, .
\end{equation}

\subsection{Quadratic Actions}

In order to construct actions, a star-integration must be defined. In
the functional space whose elements are of the form Eq. (\ref{NC_Fourier}),
any function can be integrated once the integral of a Fourier mode is
known. Hence, in the following the star-integration is introduced 

\begin{equation}
\int _{*}d^{n}x\, e_{*}^{ir\cdot x}=\left(2\pi \right)^{n}\delta \left(r\right)\: ,
\end{equation}
which yields the usual integration in the commutative limit.

Consider now the star-integral

\begin{equation}
I=\int _{*}d^{n}x\, f*g  
\end{equation}
which, in Fourier space, is written as

\begin{equation}
I=\int {d^{n}k \over \left(2\pi \right)^{n}}
{d^{n}q \over \left(2\pi \right)^{n}}\tilde{f}\left(k\right)\tilde{g}\left(q-k\right)\int _{*}d^{n}x\, 
e_{*}^{i\left[q+V\left(k,q-k\right)\right]\cdot x}\: ,
\end{equation}
following that 

\begin{equation}
I=\int {d^{n}k \over \left(2\pi \right)^{n}}d^{n}q\tilde{f}\left(k\right)\tilde{g}
\left(q-k\right)\delta \left(q+V\left(k,q-k\right)\right).
\end{equation}

Taking the structure constants nilpotent, that is, for $n>n_{*}$
\begin{equation}
\Lambda _{\quad \; \omega _{1}}^{\mu _{1}\nu }\Lambda _{\quad \quad \omega _{2}}^{\mu _{2}\omega _{1}}...
\Lambda _{\qquad \quad \omega _{n}}^{\mu _{n}\omega _{n-1}}=0\: ,
\label{nilpotency}
\end{equation}
then
\begin{equation}
\delta \left(q+V\left(k,q-k\right)\right)={\delta \left(q\right) \over \left|\det \left(\delta _{\nu }^{\mu }-
{\partial V_{\nu } \over \partial q_{\mu }}\right)\right|}=\delta \left(q\right)
\end{equation}
since $\det \left(1+M\right)=1$ if $M^{n}=0$, which holds if $\Lambda $
is nilpotent. Therefore
\begin{equation}
I=\int {d^{n}k \over \left(2\pi \right)^{n}}\tilde{f}\left(k\right)\tilde{g}\left(-k\right)=
\int d^{n}x\, f_{C}\left(x\right)g_{C}\left(x\right)\label{quadratic}
\end{equation}
where $f_{C}$, $g_{C}$ are inverse Fourier transforms using
commutative Fourier modes

\begin{equation}
f_{C}\left(x\right)=\int {d^{n}k \over \left(2\pi \right)^{n}}\tilde{f}\left(k\right)e^{ik\cdot x}.
\end{equation}
Equation (\ref{quadratic}) states that, in momentum space, quadratic
terms in the Lagrangian are the same as their commutative counterparts.
In particular, this implies that free propagators remain unchanged.

\section{Violation of Momentum Conservation}

From above considerations on can conclude that the quadratic part of a Lagrangian is not changed
and, hence, the free theory is the same as the commutative one. Thus,
the free Green function is equal to the commutative case
and the dispersion relation $\epsilon ^{2}=p^{2}+m^{2}$ is unchanged,
since it is given by the poles of the free propagator. However, one finds 
that interactions are altered by noncommutativity.

Consider a noncommutative field theory, with generic fields $A_{i}$
and an interaction term 

\begin{equation}
S_{I}=\int _{*}d^{n}x\, M_{i_{1}...i_{m}}A_{i_{1}}*...*A_{i_{m}}\: ,
\end{equation}
where $M_{i_{1}...i_{m}}$ are constants. 

In the momentum space one obtains 

\begin{equation}
S_{I}=\int \left[\prod _{i=1}^{m}{d^{n}k_{i} \over \left(2\pi \right)^{n}}\right]\tilde{M}_{i_{1}...i_{m}}
\left(\underline{k}_{m}\right)\tilde{A}_{i_{1}}\left(k_{1}\right)...\tilde{A}_{i_{m}}\left(k_{m}\right)~~,
\label{mute}
\end{equation}
where the notation $\underline{k}_{m}=\left(k_{1},...,k_{m}\right)$ has been used.
The interaction in momentum space is given by

\begin{equation}
\tilde{M}_{i_{1}...i_{m}}\left(\underline{k}_{m}\right)=M_{i_{1}...i_{m}}
\int _{*}d^{n}x\, e_{*}^{ik_{1}\cdot x}*...*e_{*}^{ik_{m}\cdot x}.
\label{new_vertex}
\end{equation}

\noindent
Notice that in Eq. (\ref{mute}) the variables $k_{i}$ are mute, so one must sum
over all $\pi $ permutations of the indices $i_{m}$:

\begin{equation}
S_{I}=\int \left[\prod _{i=1}^{m}{d^{n}k_{i} \over \left(2\pi \right)^{n}}\right]
\tilde{M}_{i_{1}...i_{m}}^{symm}\left(\underline{k}_{m}\right)\tilde{A}_{i_{1}}\left(k_{1}\right)...
\tilde{A}_{i_{m}}\left(k_{m}\right),
\end{equation}
where

\begin{equation}
\tilde{M}_{i_{1}...i_{m}}^{symm}\left(\underline{k}_{m}\right)={1 \over m!}\sum _{\pi \, perm.}
\left(-\right)^{N\left(\pi \right)}\tilde{M}_{i_{\pi \left(1\right)}...i_{\pi \left(m\right)}}\left(\underline{k}_{\pi 
\left(m\right)}\right).
\end{equation}

In order to evaluate Eq. (\ref{new_vertex}), one uses 

\begin{equation}
e_{*}^{ik_{1}\cdot x}*...*e_{*}^{ik_{m}\cdot x}=\exp _{*}\left\{ i\sum _{j=1}^{m}k_{j}\cdot x+iV^{m}
\left(\underline{k}_{m}\right)\cdot x\right\} 
\end{equation}
where 
\begin{equation}
V^{m}\left(\underline{k}_{m}\right)=V^{m-1}\left(\underline{k}_{m-1}\right)
+V\left(\sum _{i=1}^{m-1}k_{i}+V^{m-1}\left(\underline{k}_{m-1}\right),k_{m}\right)
\end{equation}
with $V^{2}\left(\underline{k}_{2}\right)=V\left(k_{1},k_{2}\right)$.
This yields both the noncommutative energy-momentum law
and the noncommutative vertex

\begin{equation}
\tilde{M}_{i_{1}...i_{m}}\left(\underline{k}_{m}\right)
=\left(2\pi \right)^{n}\delta \left(\sum _{i=1}^{m}k_{i}+V^{m}\left(\underline{k}_{m}\right)\right)M_{i_{1}...i_{m}}.
\label{nc_vertex}
\end{equation}

\noindent
Hence, the noncommutative energy-momentum law for the full theory $\tilde{M}_{i_{1}...i_{m}}^{symm}$ vertex reads

\begin{equation}
\sum _{i=1}^{m}k_{i}+V^{m}\left(\underline{k}_{\pi \left(m\right)}\right)=0
\label{cons_momentum}
\end{equation}
for all $m!$ permutations of indices, $\pi $. 

Thus, it can be seen that the energy-momentum conservation is violated
as the theory is not invariant under translations. Indeed, in a translation $x^{\mu }\rightarrow x^{\mu }+b^{\mu }$,
the commutator of the coordinates is changed by

\begin{equation}
\left[x^{\mu },x^{\nu }\right]\rightarrow i\, \Lambda _{\quad \omega }^{\mu \nu }x^{\omega }+i\, \theta ^{\mu \nu }\: ,
\end{equation}
that is, a constant term $\theta ^{\mu \nu }=\Lambda _{\quad \omega }^{\mu \nu }b^{\omega }$
is added to the commutator of the coordinates. So, the interaction
vertex becomes

\begin{equation}
 \tilde{M}_{i_{1}...i_{m}}\left(\underline{k}_{m}\right)\rightarrow 
 \left(2\pi \right)^{n}\delta \left(\sum _{i=1}^{m}k_{i}+V^{m}
\left(\underline{k}_{\pi \left(m\right)}\right)\right)M_{i_{1}...i_{m}}\exp 
\left\{ i\theta ^{m}\left(\underline{k}_{m}\right)\right\} \nonumber 
\end{equation}
where

\begin{equation}
\theta ^{m}\left(\underline{k}_{m}\right)=\theta ^{m-1}\left(\underline{k}_{m-1}\right)+\theta 
\left(\sum _{i=1}^{m-1}k_{i}+V^{m-1}\left(\underline{k}_{m-1}\right),k_{m}\right)
\end{equation}
and $\theta ^{2}\left(\underline{k}_{2}\right)=k_{1\mu }\theta ^{\mu \nu }k_{2\nu }$.
Therefore, the interaction vertex is changed by an overall oscillating
momentum-dependent factor which breaks invariance under translations. 
This shows that translations give always rise to
a constant term in the noncommutative tensor.

\section{Kinematic Applications}

\subsection{Preliminaries}

The first non-trivial consequence arising from the new interaction vertex
takes place when considering three particles. The energy-momentum equation
is found to be

\begin{equation}
k_{1}+k_{2}+k_{3}+V\left(k_{1},k_{2}\right)+V\left(k_{1}+k_{2}
+V\left(k_{1},k_{2}\right),k_{3}\right)=0
\label{cons_decay}
\end{equation}
and similar expressions for all permutations of the indices.

Let us first consider the model

\begin{equation}
\Lambda _{\quad \omega _{1}}^{\mu _{1}\nu }\Lambda _{\quad \omega _{2}}^{\mu _{2}\omega _{1}}=0~~,
\label{nil2}
\end{equation}
which is consistent with the Jacobi identity Eq. (\ref{Jacobi}).

The energy-momentum equation reads
\begin{equation}
k_{1}+k_{2}+k_{3}+V\left(k_{1},k_{2}\right)=0\: ,
\label{cons_decay_simple}
\end{equation}
where
\begin{equation}
V_{\omega }\left(k_{1},k_{2}\right)={1 \over 2}k_{1\mu }k_{2\nu }\Lambda _{\quad \omega }^{\mu \nu }\; .
\end{equation}

\noindent
Eq. (\ref{nil2}) admits non-trivial covariant solutions. For
instance, consider a constant antisymmetric tensor $\Lambda ^{\mu \nu }=-\Lambda ^{\nu \mu }$
with non-trivial kernel, $\det \Lambda =0$, and a non-vanishing
vector $r^{\omega }$ belonging to this kernel. Hence a solution is
given by
\begin{equation}
\Lambda ^{\mu \nu \omega }=\Lambda ^{\mu \nu }r^{\omega }\: .
\label{solution}
\end{equation}
In four dimensions one can parametrize $\Lambda ^{\mu \nu }$ with
two spatial vectors $\vec{E}$ and $\vec{B}$
\begin{equation}
\Lambda ^{\mu \nu }=\left(\begin{array}{cccc}
 0 & E_{x} & E_{y} & E_{z}\\
 -E_{x} & 0 & -B_{z} & B_{y}\\
 -E_{y} & B_{z} & 0 & -B_{x}\\
 -E_{z} & -B_{y} & B_{x} & 0\end{array}\right)\quad ,\quad r_{\nu }=\left(\begin{array}{c}
 r_{0}\\
 r_{x}\\
 r_{y}\\
 r_{z}\end{array}\right)\: .\end{equation}
 Condition Eq. (\ref{solution}) implies that
\begin{equation}
r^{2}=\left|\vec{r}\right|^{2}\left[\left({B \over E}\sin \delta \right)^{2}-1\right],
\label{tachyon?}
\end{equation}
with $\delta $ being the angle between $\vec{B}$ and $\vec{r}$.
The massless, massive and tachyon regimes of $V$ can be easily identified.
Assuming that $\Lambda ^{\mu \nu }$ is a Lorentz tensor, there
are always inertial frames where $\vec{E}$ is non-vanishing, and
the above expression holds only for such frames. If $B<E$ (a Lorentz-invariant
inequality) then $r^{\omega }$ behaves like a tachyon; otherwise,
the behaviour of $r^{\omega }$ will depend on $\delta $. 

From the momentum expression, Eq. (\ref{cons_decay_simple}), one
gets
\begin{equation}
\Lambda ^{\mu \nu }\left(k_{1}+k_{2}+k_{3}\right)_{\nu }=0\: ,
\label{kernel_cons}
\end{equation}
which states that the vector sum of the momenta belongs to the non-trivial
kernel of the noncommutative tensor. There follows the expressions

\begin{equation}
\Lambda ^{\mu \nu }k_{\nu }=\left(\begin{array}{c}
 \vec{E}\cdot \vec{k}\\
 -k_{0}\vec{E}+\vec{B}\times \vec{k}\end{array}\right)
\label{lambda_k}
\end{equation}
and
\begin{equation}
q_{\mu }\Lambda ^{\mu \nu }k_{\nu }=\vec{E}\cdot \left(q_{0}\vec{k}-k_{0}\vec{q}\right)
+\vec{B}\cdot \left(\vec{k}\times \vec{q}\right)\: .
\end{equation}

Eqs. (\ref{kernel_cons}) and (\ref{lambda_k}) imply that the three-momentum
is conserved along the direction of $\vec{E}$. Energy is conserved
if the total three-momentum $\sum \vec{k_{i}}$ lies along the direction
of $\vec{B}$. Also,

\begin{equation}
k_{1}\Lambda k_{2}=-k_{1}\Lambda k_{3}=k_{2}\Lambda k_{3}
\end{equation}
and one is required only to study Eq. (\ref{cons_decay_simple}) with
$V\left(k_{1},k_{2}\right)$ and $-V\left(k_{1},k_{2}\right)$. 
Notice
that the second case is obtained by performing 
the transformation $\vec{E},\vec{B}\rightarrow -\vec{E},-\vec{B}$.

\subsection{The GZK cutoff}

Let us discuss the GZK cutoff in the context of our 
noncommutative model, considering the dominant resonance

\begin{equation}
p+\gamma _{CMB}\rightarrow \Delta _{1232}\: .
\label{GZK}
\end{equation}

A violation of the GZK cutoff can arise in the
context of the model

\begin{equation}
\Lambda _{\quad \omega _{1}}^{\mu _{1}\nu }
\Lambda _{\qquad \omega _{2}}^{\mu _{2}\omega _{1}}\Lambda _{\qquad \omega _{3}}^{\mu _{3}\omega _{2}}=0\; ,
\end{equation}
with $\Lambda _{\quad \omega _{1}}^{\mu _{1}\nu }\Lambda _{\quad \omega _{2}}^{\mu _{2}\omega _{1}}\neq 0$.
This cannot be implemented by model Eq. (\ref{solution}), as the analysis is fairly complicate. 
The equation for the momentum is given by

\begin{equation}
\left(k_{1}+k_{2}+k_{3}\right)_{\omega }=k_{1\mu }k_{2\nu }\Lambda _{\quad \lambda }^{\mu \nu }
\left[-{1 \over 2}\delta _{\omega }^{\lambda }+{\left(k_{1}-k_{2}\right)_{\alpha } \over 12}
\Lambda _{\quad \omega }^{\alpha \lambda }\right]\label{GZK_v2}
\end{equation}
where Eq. (\ref{cons_decay}) has been recursively used as well as the fact
that the cubic terms in $\Lambda$ vanish.

Condition Eq. (\ref{GZK_v2}) can be replaced by a simpler one, more suitable for
phenomenological considerations.
Dropping quadratic terms in the momentum and taking
into account that the proton has the highest energy and the $\Delta $
the second highest energy, one can write the new momentum 
equation for the reaction (\ref{GZK}) as

\begin{equation}
\left(k_{p}+k_{\gamma }\right)^{\mu }=k_{\Delta }^{\mu }-s^{\mu }{\epsilon _{p}^{2} \over M^{2}}\epsilon _{\Delta }~~,
\label{GZK_funciona}
\end{equation}
where the dimensionless vector $s^{\mu }$ is of the order of unity and
$M$ is the typical noncommutative mass scale. In this case, the process
is impossible if $s^{0}>0$ and $\epsilon _{p}>M$, which sets the
scale of noncommutativity.

It is not difficult to see that a cubic term
in the dispersion equation\cite{Amelino,Konopka,Bertolami10} can explain the violation of this
cutoff, even though general arguments, based on coordinate invariance and causality, 
may prevent this term if 4-momentum is conserved\cite{Lehnert2}. These objections do not apply to our proposal 
given the breaking of translational invariance.  
In fact, Eq. (\ref{GZK_funciona}) can be obtained postulating a new equation of
dispersion by the substitution 

\begin{equation}
k^{\mu }\rightarrow k^{\mu }+s^{\mu }{\epsilon ^{2} \over M^{2}}\lambda 
\end{equation}
where $\lambda$ represents the typical energy of the product of
the reaction. This leads to the following dispersion relation
\begin{equation}
m^{2}=\epsilon ^{2}-p^{2}+2s^{\mu }v_{\mu }{\lambda \over M^{2}}\epsilon^{3}~~,
\label{cubic_eq_disp}
\end{equation}
where $v^{\mu }=\left(1,\vec{v}\right)$ is the four-vector velocity,
which is assumed to be ultra relativistic. Only the lower order terms
of the correction were kept.
Thus, it is as if a cubic term had been introduced into the dispersion 
relation and the GZK cutoff is evaded if $M \simeq 4\times 10^{19}eV$. 

%We point out that
%this model differs from other approaches\cite{Amelino} in
%the sense that Eq. (\ref{cubic_eq_disp}) depends on the
%energy of the product of the reaction as well as on the geometry of the process% in
%question. Clearly, a sensible choice for M is $M \simeq 4\times 10^{19}eV$. 

\section{Conclusions}

In this work we have discussed a noncommutative field theory where
the coordinates have a Lie-algebra commutator as Eq. (\ref{intro}) with 
nilpotent structure constants.
This breaks Lorentz symmetry as well as translational
invariance. Free theory is unchanged so the propagators and the dispersion
relations are not modified. Interaction however, lead to a new energy-momentum law, 
which follows from the breaking of translational invariance. The
kinematics of such law was established and considered as a possible explanation 
for the violation of the GZK
cutoff, if one chooses the noncommutative mass scale at $M \simeq 4\times 10^{19}eV$.
The use of the present results to the
other astrophysical puzzles will be considered elsewhere.

\end{document}